\documentclass[aps,nofootinbib,twocolumn,prd,eqsecnum,showpacs,showkeys,preprintnumbers]{revtex4-1}

\usepackage[caption=false]{subfig}
\usepackage{graphicx}
\usepackage{amsmath}
\usepackage{amsfonts}
\usepackage{amssymb}
\usepackage{color}
\usepackage{bm}
\usepackage{mathrsfs}
\usepackage{epstopdf}
\usepackage{url}
\usepackage{footnote}
\usepackage{textcomp}
\usepackage{dsfont}
\usepackage{ulem}
\usepackage{hyperref}

\makeatletter
\newcommand*{\rom}[1]{\expandafter\@slowromancap\romannumeral #1@}
\makeatother

\begin{document}

\title{Quasi-normal modes of massless scalar fields for charged black holes in the Palatini-type gravity}
\author{Che-Yu Chen $^{1,2}$}
\email{b97202056@gmail.com}
\author{Pisin Chen $^{1,2,3}$}
\email{pisinchen@phys.ntu.edu.tw}
\date{\today}

\affiliation{
${}^1$Department of Physics and Center for Theoretical Sciences, National Taiwan University, Taipei, Taiwan 10617\\
${}^2$LeCosPA, National Taiwan University, Taipei, Taiwan 10617\\
${}^3$Kavli Institute for Particle Astrophysics and Cosmology, SLAC National Accelerator Laboratory, Stanford University, Stanford, CA 94305, U.S.A.
}

\begin{abstract}
The scrutiny of black hole perturbations and its application to testing gravity theories have been a very crucial frontier in modern physics. In this paper we study the quasi-normal modes (QNM) of massless scalar perturbations for various charged black holes in the Palatini-type theories of gravity. Specifically, we consider the Palatini $f(R)$ theory coupled with Born-Infeld nonlinear electrodynamics and the Eddington-inspired-Born-Infeld gravity (EiBI) coupled with Maxwell electromagnetic fields. Special attention is paid to Einstein-Born-Infeld black holes, EiBI charged black holes and Born-Infeld charged black holes within $R\pm R^2$ gravity. These charged black holes are shown to be stable and their quasi-normal frequencies, including the frequencies in the eikonal limit, are calculated by using the Wentzel-Kramers-Brillouin (WKB) method up to the 6th order. We compare the results with the Reissner-Nordstr\"om charged black hole and prove that both the changes in the gravity sector and the matter sector of the action alter the QNM spectra.
\end{abstract}


\maketitle

\section{Introduction}
It is generally agreed that Einstein's general relativity (GR) is not a complete theory in the sense that it inevitably predicts the existence of spacetime singularities, such as the big bang singularity and the black hole singularity. These singularities are expected to be nonphysical and it is hoped that quantum effects may ameliorate them in the extremely high curvature limit. Unfortunately, how these quantum effects would be incorporated with GR is still an open question. In view of this, it is phenomenologically acceptable to consider some classical extended theories of gravity as an effective theory of a more fundamental yet unknown quantum theory of gravity \cite{Capozziello:2011et}. An important issue naturally arises: is there a way to falsify them through the observational data? 

In recent years, one of the most fascinating events in physics is the direct detection of gravitational waves from binary black hole mergers by LIGO \cite{Abbott:2016blz,Abbott:2017oio}. These observations not only confirm the correctness of the predictions of GR, but also provide a perfect arena to test other alternative gravitational theories, taking advantage of the strong spacetime curvature of the sources. Not quite long after the first detection of gravitational waves, the LIGO-VIRGO collaboration even managed to detect the gravitational waves emitted from the merger of binary neutron stars \cite{TheLIGOScientific:2017qsa}. The most important meaning of this event is that prompt observations of electromagnetic signals emitted from the sources are also attainable. The direct detection of gravitational waves and their accompanied electromagnetic signals may not only help us to uncover the hidden properties of the sources, but also to constrain gravitational theories, such as via the speed of gravitational waves \cite{Baker:2017hug,Lombriser:2015sxa,Lombriser:2016yzn,Sakstein:2017xjx}.

The observations of gravitational waves from merger events usually involve the analysis of the ringing signals. This stage of signals is called \textit{ringdown} and it corresponds to the final stage of a merger event. In theory, the ringdown stage can be well-described by the perturbations of the black hole which is the product of the merger, during which the frequencies of the gravitational waves are characterized by quasi-normal modes (QNM) with complex numbers. The real part of the frequency describes the oscillations of the perturbations and the imaginary part corresponds to the decay of the amplitude, as long as the black hole is stable. Furthermore, these QNM frequencies only depend on the parameters characterizing the black holes, such as the mass, the charge and the spin. If the underlying gravitational theory contains some additional parameters, these QNM frequencies would naturally depend on them as well. That is the major reason why investigating the QNMs ringing may shed some light on constraining the extended theories of gravity. Furthermore, probing signatures of the black hole phase transitions in modified gravities via QNM frequencies has also been shown to be possible \cite{Mahapatra:2016dae}. See Refs~\cite{Nollert:1999ji,Berti:2009kk,Konoplya:2011qq,Berti:2015itd} for reviews on the current progress of the field. 

In this paper, we will focus on the massless scalar perturbations of the charged black holes within two Palatini-type gravity theories: the Palatini $f(R)$ gravity coupled with Born-Infeld nonlinear electrodynamics (NED) and the Eddington-inspired-Born-Infeld (EiBI) gravity coupled with linear electromagnetic fields. The QNM frequencies, including the QNMs in the eikonal limit, will be calculated using the WKB method up to the 6th order \cite{Schutz:1985zz,Iyer:1986np,Konoplya:2003ii,Matyjasek:2017psv}.{\footnote{The extension of the WKB method to the 13th order does improve the accuracy of the QNM frequencies \cite{Matyjasek:2017psv}. However, using the 6th order WKB method has been rather sufficient to see the overall tendency of QNM frequencies affected by modifications in the gravitational action considered in this paper.}} The comparison of the frequencies with those for the standard Reissner-Nordstr\"om (RN) black hole will be exhibited. 

The solutions of charged black holes within the Palatini $f(R)$ gravity coupled with Born-Infeld NED have been studied in Ref.~\cite{Olmo:2011ja}. Due to the non-linear Born-Infeld modifications and the $f(R)$ corrections in the gravitational sector, the black hole structure, especially at small radius, would be significantly changed. For example, there could be single, double, or even zero event horizons, depending on the choice of the parameter space. It has also been shown that in a $R+\alpha R^2$ gravity with a negative $\alpha$, the black hole singularity is replaced with a finite size wormhole structure \cite{Bambi:2015zch}. Furthermore, the well-known Einstein-Born-Infeld black hole can be obtained by choosing $f(R)=R$ in this theory. The metric functions and the properties of this black hole have been widely studied in the literature \cite{Breton:2001yk,Dey:2004yt,Cai:2004eh,Fernando:2003tz}. The interior structure of the black hole can also be rather different from that of the standard RN black hole. For instance, the intensity of the divergence of the curvature scalar would decrease for a Einstein-Born-Infeld black hole \cite{Olmo:2011ja}.

 The EiBI theory was firstly proposed in Ref.~\cite{Banados:2010ix}, where the big bang singularity in the early universe has been shown to be avoidable \cite{Banados:2010ix,Scargill:2012kg}. The properties of charged black holes are also studied in Refs.~\cite{Wei:2014dka,Sotani:2014lua} (see Ref.~\cite{BeltranJimenez:2017doy} for a recent review on the EiBI gravity). The black hole structure in the EiBI theory would be drastically different from the standard RN black hole due to the Born-Infeld modifications in the gravity sector. We expect that these deviations would be manifested in the QNM frequencies. Another interesting issue is to compare the difference in QNM frequencies between the EiBI charged black holes and the Einstein-Born-Infeld black holes. One can then distinguish contributions to the QNM frequencies between that from the gravity sector (EiBI) and that from the matter sector (Einstein-BI). This is the main goal of this work.

There are actually two main motivations to study QNMs of the charged black holes, even though an astrophysical charged black hole usually neutralizes itself rather quickly. First, the eventual black hole after the neutron star merger should initially have charges due to the existence of some high energy particles during the merge \cite{TheLIGOScientific:2017qsa}. The ringing time scale of the gravitational wave is of the order of the event horizon size of the black hole. However, the time scale of the neutralization should be of the order of the radius inside which the matter surrounds the black hole \cite{Franklin:2011qn}. Therefore, the neutralization time scale should be longer than the ringing time, and the initial ringdown signals of the black hole in the binary neutron star mergers should in principle contain the imprints of the charges. Secondly, the Palatini-type gravity theories, in most cases, simply reduce to GR with an effective cosmological constant in the absence of matter. This can be seen from the nature of the field equations and we will explain it in more details later.    
 
This paper is outlined as follows. In section \ref{sectII}, the equation describing the massless scalar perturbations of a static and spherically symmetric black hole within the Palatini-type gravity theories is presented.  In sections \ref{sectIII} and \ref{sectIV}, we review the derivation of the metric functions of charged black holes in the Palatini $f(R)$ gravity coupled with Born-Infeld NED and in the EiBI gravity coupled with Maxwell fields, respectively. In section \ref{sectV}, we calculate the QNM frequencies of these charged black holes with a WKB technique. The frequencies in the eikonal regime are also presented. We finally conclude in section \ref{conclu}.

\section{Massless scalar perturbations of black holes in the Palatini-type gravity theories}\label{sectII}

In a modified theory of gravity constructed upon the Palatini variational principle, the physical metric $g_{\mu\nu}$ and the affine connection $\Gamma$ are regarded as independent variables. However, the matter Lagrangian is still assumed to be coupled with the physical metric $g_{\mu\nu}$ only, as in GR. Therefore, the matter fields would follow the geodesics defined by this metric and the conservation equation of the energy momentum tensor follows the standard form $\nabla_\nu T^{\mu\nu}=0$, where the covariant derivative $\nabla_\nu$ is defined solely by the physical metric \cite{Koivisto:2005yk}. 

In this regard, if we consider the scalar perturbation from a massless scalar field $\Phi$ around a black hole background, the equation of motion of this perturbation is the Klein-Gordon equation
\begin{equation}
\Box\Phi=\frac{1}{\sqrt{-g}}\partial_\alpha(g^{\alpha\beta}\sqrt{-g}\partial_\beta\Phi)=0.
\label{KG}
\end{equation}
It can be seen that the equation directly depends on the metric $g_{\mu\nu}$ describing the black hole spacetime. For the black hole spacetime, we introduce the most general form of a static and spherically symmetric metric
\begin{equation}
ds_g^2=-\psi^2(r)\bar{f}(r)dt^2+\frac{1}{\bar{f}(r)}dr^2+r^2d\Omega^2.
\label{gmetric}
\end{equation}
After decomposing the scalar field $\Phi$ as follows:
\begin{equation}
\Phi(t,r,\theta,\phi)=\sum_{l,m}\frac{1}{r}\psi_l(t,r)Y_{lm}(\theta,\phi),
\end{equation}
the Klein-Gordon equation \eqref{KG} can be written as
\begin{equation}
-\partial_t^2\psi_l+\partial_{r_*}^2\psi_l=V(r)\psi_l,
\label{KGeq}
\end{equation}
where 
\begin{equation}
V(r)=\psi^2\bar{f}\Big[\frac{l(l+1)}{r^2}+\frac{1}{r\psi}\frac{d}{dr}(\bar{f}\psi)\Big],
\end{equation}
and $r_*$ refers to a generalized tortoise radius defined by
\begin{equation}
\frac{dr_*}{dr}=\frac{1}{\psi \bar{f}}.
\end{equation}
Using the Fourier decomposition ($\partial_{t}\rightarrow -i\omega$), Eq.~\eqref{KGeq} can be further written as
\begin{equation}
\partial_{r_*}^2\psi_l+\Big[\omega^2-V(r)\Big]\psi_l=0.
\label{perturbationeq}
\end{equation} 

It is worth emphasizing that the equation for the scalar field perturbations can be written in a Schr\"{o}dinger-like form even if the metric function $\psi(r)$ could be any function of $r$. One should as well keep in mind that the divergencelessness of the energy momentum tensor $\nabla_\nu T^{\mu\nu}=0$ plays a central role of the derivation of Eq.~\eqref{perturbationeq}. In a more general metric-affine theory in which the couplings between the matter sector and the affine connection, or non-minimal couplings between matter and gravity are introduced, the $g$-divergencelessness of $T^{\mu\nu}$ breaks down and Eq.~\eqref{perturbationeq} is not valid anymore \cite{Koivisto:2005yk,Harko:2010hw}.

\section{Charged black holes in Palatini $f(R)$ gravity with nonlinear electrodynamics}\label{sectIII}
In this section, we will firstly consider the Palatini $f(R)$ gravity coupled with NED. The derivation of the field equations and the metric functions of the black holes are clearly elucidated in Ref.~\cite{Olmo:2011ja}. Here we will mainly follow the approach in \cite{Olmo:2011ja} while recast the final expressions of the metric functions in a more suitable form to calculate the QNMs. The reason of choosing NED instead of the standard Maxwell electromagnetic fields will become clearer later. 

The action of the Palatini $f(R)$ gravity coupled with NED reads
\begin{equation}
\mathcal{S}_{1}=\frac{1}{16\pi}\int d^4x\sqrt{-g}f(R)+\frac{1}{8\pi}\int d^4x\sqrt{-g}\phi(X,Y),
\label{action1}
\end{equation}
where we have set $G=c=1$. In the matter sector, $\phi(X,Y)$ is a function of gauge field invariants defined by \cite{Olmo:2011ja}
\begin{equation}
X\equiv-\frac{1}{2}F_{\mu\nu}F^{\mu\nu}\,,\qquad Y\equiv-\frac{1}{2}F_{\mu\nu}F^{*\mu\nu},
\end{equation}
where $F^{*\mu\nu}\equiv\frac{1}{2}\epsilon^{\mu\nu\alpha\beta}F_{\alpha\beta}$ is the dual of the field strength. The standard Maxwell electromagnetic fields are recovered when $\phi(X,Y)=X$. For the sake of simplicity, we will assume a vanishing magnetic field, i.e., $Y=0$, in the rest of this paper.

\subsection{Equations of motion}
Because the theory is based on the Palatini variational principle, the affine connection and the physical metric are independent variables. To derive the field equations, we need to vary the action with respect to them separately. The variation of the action \eqref{action1} with respect to $g_{\mu\nu}$ leads to
\begin{equation}
f_R R_{(\mu\nu)}(\Gamma)-\frac{1}{2}g_{\mu\nu}f(R)=8\pi T_{\mu\nu},
\label{eq1g}
\end{equation}
where $f_R\equiv df/dR$ and $T_{\mu\nu}$ stands for the energy momentum tensor of NED:
\begin{equation}
T_{\mu\nu}=\frac{1}{4\pi}\Big(\phi_XF_{\mu\sigma}{F_{\nu}}^\sigma+\frac{1}{2}\phi g_{\mu\nu}\Big),
\label{nedTmunu}
\end{equation}
where $\phi_X\equiv d\phi/dX$. By taking a trace of Eq.~\eqref{eq1g}, we have
\begin{align}
f_RR-2f&=8\pi T\nonumber\\
&=4(\phi-X\phi_X).\label{traceeq1g}
\end{align}
Since the trace of energy momentum tensor is zero for standard Maxwell fields, i.e., $T=0$ if $\phi=X$, Eq.~\eqref{traceeq1g} is an algebraic equation for $R$ and $R$ eventually turns out to be a constant $R_0$. In this case, the metric function reduces to the RN-AdS/dS type whose QNMs have been widely studied in the literature \cite{Wang:2000gsa,Wang:2000dt,Mellor:1989ac,Molina:2003ff,Jing:2003wq}. To obtain new dynamics characterizing the modifications in the action, we consider the matter source whose $T\neq 0$. That is why we include NED rather than standard Maxwell fields in this work. 

On the other hand, the conservation equation of NED reads
\begin{equation}
\frac{1}{\sqrt{-g}}\partial_{\mu}(\sqrt{-g}\phi_XF^{\mu\nu})=0.
\label{conser}
\end{equation}
For the spacetime metric \eqref{gmetric} and a purely radial electric field, i.e., $A_{\mu}=(A_0(r),0,0,0)$, Eq.~\eqref{conser} can be solved to obtain
\begin{equation}
\phi_X^2X=\frac{Q^2}{r^4},
\label{21}
\end{equation}
where $Q$ is an integration constant and it will be regarded as the charge of the black hole in the end \cite{Olmo:2011ja}. 

Furthermore, the variation of the action \eqref{action1} with respect to the affine connection $\Gamma$ results in
\begin{equation}
\nabla^\Gamma_{\lambda}(\sqrt{-h}h^{\mu\nu})=0,\label{covh}
\end{equation}
where the auxiliary metric $h_{\mu\nu}$ is defined by a conformal transformation of the physical metric: $h_{\mu\nu}\equiv f_Rg_{\mu\nu}$. The covariant derivative $\nabla^\Gamma_{\lambda}$ is defined by the affine connection, and $h^{\mu\nu}$ is the inverse of $h_{\mu\nu}$. This equation implies that the affine connection can be regarded as the Levi-Civita connection of the auxiliary metric $h_{\mu\nu}$ and it gives the second order differential equation with which the metric functions can be obtained. 

In general, we can express the line element of the auxiliary metric $h_{\mu\nu}$ as follows:
\begin{equation}
ds^2_h=-G^2(x)F(x)dt^2+\frac{1}{F(x)}dx^2+x^2d\Omega^2.
\label{hmetric}
\end{equation}
Since $g_{\mu\nu}$ and $h_{\mu\nu}$ are conformally related, we immediately get following identities:
\begin{equation}
G^2(x)F(x)=f_R\psi^2(r)\bar{f}(r)\,,\quad\Big(\frac{dx}{dr}\Big)^2=f_R\frac{F(x)}{\bar{f}(r)}\,,\label{identity1}
\end{equation}
and 
\begin{equation}
x^2=f_Rr^2.\label{identity2}
\end{equation}
Using the metric ansatz \eqref{hmetric} and solving Eqs.~\eqref{eq1g} and \eqref{covh}, we have
\begin{equation}
G(x)=\textrm{constant},
\end{equation}
which the constant can be set to unity after a time rescaling. Furthermore, the function $F(x)$ can be expressed as
\begin{equation}
F(x)=1-\frac{2M(x)}{x},
\end{equation}
where the mass function $M(x)$ satisfies
\begin{equation}
\frac{dM(x)}{dx}=\frac{x^2}{4f_R^2}\Big(f+2\phi\Big),
\end{equation}
and \cite{Olmo:2011ja}
\begin{equation}
\frac{dM(r)}{dr}=\frac{r^2}{4f_R^{3/2}}(f+2\phi)\Big(f_R+\frac{r}{2}f_{R,r}\Big).
\label{Mdr}
\end{equation}

\subsection{Born-Infeld NED and metric functions}
From now on, we will consider a specific form of $\phi$: the Born-Infeld NED:
\begin{equation}
\phi(X,Y)=2\beta_m^2\Big(1-\sqrt{1-\frac{X}{\beta_m^2}-\frac{Y^2}{4\beta_m^4}}\Big).
\end{equation}
For the sake of later convenience, we apply the following dimensionless rescalings:
\begin{equation}
\frac{Q}{r_s}\rightarrow Q\qquad \beta_mr_s\rightarrow \beta_m\qquad\frac{r}{r_s}\rightarrow r\,,
\end{equation} 
where $r_s/2\equiv M_0$ denotes the mass of the black hole seen by an observer infinitely away. Within the framework of Born-Infeld NED, Eq.~\eqref{21} leads to
\begin{equation}
\frac{r_s^2X}{\beta_m^2}=\frac{Q^2}{Q^2+\beta_m^2r^4}.\label{3.8}
\end{equation}
Using Eq.~\eqref{3.8}, we can obtain
\begin{align}
r_s^2\phi&=\beta_m^2\Big(1-\sqrt{\frac{z^4}{1+z^4}}\Big)\equiv\beta_m^2\tilde{\phi}\,,\\
\phi_X&=\sqrt{1+\frac{Q^2}{\beta_m^2r^4}}\equiv\sqrt{1+\frac{1}{z^4}}\,,
\end{align}
where we have defined a dimensionless radius $z\equiv r\sqrt{\beta_m/Q}\equiv r/r_m$. Furthermore, the trace of the energy momentum tensor can be written as
\begin{align}
r_s^2T&=\frac{r_s^2}{2\pi}(\phi-X\phi_X)\nonumber\\
&=\frac{-\beta_m^2}{2\pi\sqrt{1+\frac{1}{z^4}}}\Big[\frac{1}{z^4}+2\Big(1-\sqrt{1+\frac{1}{z^4}}\Big)\Big]\nonumber\\
&\equiv\beta_m^2\tilde{T}.
\label{trTd}
\end{align}
Note that $\tilde{\phi}$ and $\tilde{T}$ only contain the parameter $z$.

The differential equation of the mass function \eqref{Mdr} can be rewritten in terms of $z$:
\begin{align}
\frac{dM}{dz}&=\frac{r_m^3z^2r_s^3}{4f_R^{3/2}}(f+2\phi)\Big(f_R+\frac{z}{2}f_{R,z}\Big)\nonumber\\
&=(r_m^3\beta_m^2r_s)\frac{z^2}{4f_R^{3/2}}(\tilde{f}+2\tilde{\phi})\Big(f_R+\frac{z}{2}f_{R,z}\Big),\label{massz}
\end{align}
where $\tilde{f}$ is defined by $r_s^2f\equiv\beta_m^2\tilde{f}$. In general, $\tilde{f}$ is dimensionless and it can be written in terms of $\tilde{T}$ via Eqs.~\eqref{traceeq1g} and \eqref{trTd}. For example, in GR we have $f=R$ and $\tilde{f}=-8\pi\tilde{T}$.

To proceed, we integrate Eq.~\eqref{massz} to derive the mass function
\begin{align}
M(z)&=M_0-\frac{r_m^3\beta_m^2r_s}{2}\int_z^\infty\frac{z^2}{2f_R^{3/2}}(\tilde{f}+2\tilde{\phi})\Big(f_R+\frac{z}{2}f_{R,z}\Big)dz\nonumber\\
&\equiv M_0+\frac{r_m^3\beta_m^2r_s}{2}\tilde{G}(z)\nonumber\\
&=M_0[1+r_m^3\beta_m^2\tilde{G}(z)].
\end{align}
According to the identities \eqref{identity1} and \eqref{identity2}, the metric functions can be obtained as follows \cite{Olmo:2011ja}
\begin{equation}
\psi^2\bar{f}=\frac{1}{f_R}\Big[1-\frac{1+r_m^3\beta_m^2\tilde{G}(z)}{f_R^{1/2}r_mz}\Big],\label{preintegral}
\end{equation}
and
\begin{equation}
\bar{f}=\Big[\frac{d(zf_R^{1/2})}{dz}\Big]^{-2}f_R\Big[1-\frac{1+r_m^3\beta_m^2\tilde{G}(z)}{f_R^{1/2}r_mz}\Big].\label{integral}
\end{equation}
\subsection{Einstein-Born-Infled charged black holes}
If $f(R)=R$, the black hole solution reduces to the Einstein-Born-Infeld black hole. The metric functions can be obtained via Eqs.~\eqref{preintegral} and \eqref{integral} \cite{integrals,integrals2}: 
\begin{widetext}
\begin{align}
\psi(r)&=1,\nonumber\\
\bar{f}(r)&=1-\frac{1}{r}-\frac{2\beta_m^2}{3}\Big[\sqrt{r^4+r_m^4}-r^2-\frac{r_m^3}{r}F\Big(\cos^{-1}{\frac{r^2-r_m^2}{r^2+r_m^2}\,,\frac{1}{\sqrt{2}}}\Big)\Big]\nonumber\\
&=1-\frac{1}{r}-\frac{2\beta_m^2}{3}\Big[\sqrt{r^4+r_m^4}-r^2-\frac{2r_m^4}{r^2}F\Big(\frac{1}{4},\frac{1}{2},\frac{5}{4},-\frac{r_m^4}{r^4}\Big)\Big].\label{EBIpsif}
\end{align}
\end{widetext}
The result is consistent with that presented in Refs.~\cite{Breton:2001yk,Dey:2004yt,Cai:2004eh,Fernando:2003tz}. Note that the elliptic function of first kind can be expressed as the Hypergeometric function by
\begin{equation}
F\Big(\cos^{-1}{\frac{z^2-1}{z^2+1}\,,\frac{1}{\sqrt{2}}}\Big)=\frac{2}{z}F\Big(\frac{1}{4},\frac{1}{2},\frac{5}{4},-\frac{1}{z^4}\Big).
\end{equation}
At large radius, or small $Q/\beta_m$, the metric function $\bar{f}$ can be approximated as follows
\begin{equation}
\bar{f}=1-\frac{1}{r}+\frac{Q^2}{r^2}-\frac{Q^4}{20\beta_m^2r^6}+\frac{Q^6}{72\beta_m^4r^{10}}+\mathcal{O}(r^{-12})\,.\label{329}
\end{equation}
It can be seen that at this limit ($z\gg1$), the solution reduces to the RN black hole.  

\subsection{Another example: $f(R)=R+\alpha R^2$}
As the simplest generalization of gravitational part, we consider $f(R)=R+\alpha R^2$. In this case, we have $f_R=1+2\alpha R$ and $R=-8\pi T=-8\pi\beta_m^2\tilde{T}/r_s^2$. Therefore, we get
\begin{equation}
f_R=1-\frac{16\pi\alpha\beta_m^2}{r_s^2}\tilde{T}\equiv1-2\gamma\tilde{T}\,,\quad \tilde{f}=8\pi(-\tilde{T}+\gamma\tilde{T}^2),
\end{equation}
where the dimensionless parameter $\gamma\equiv8\pi\alpha\beta_m^2/r_s^2$ quantifies the correction from the additional $R^2$ term in the gravitational sector. 

The exact expression of metric functions is not available in this case. However, we can still derive the approximated solution at large $z$ limit ($r\gg r_m$). After normalizing the parameter $\alpha$ as follows: $\alpha/r_s^2\rightarrow\alpha$, the approximated metric functions at large $z$ are
\begin{widetext}
\begin{align}
\psi^2\bar{f}&=1-\frac{1}{r}+\frac{Q^2}{r^2}-\frac{Q^4}{20\beta_m^2r^6}-\frac{2\alpha Q^4}{\beta_m^2r^8}+\frac{3\alpha Q^4}{\beta_m^2r^9}+\Big(\frac{1}{72}-4\alpha\beta_m^2\Big)\frac{Q^6}{\beta_m^4r^{10}}+\mathcal{O}(r^{-12})\,,\nonumber\\
\bar{f}&=1-\frac{1}{r}+\frac{Q^2}{r^2}-\frac{Q^4}{20\beta_m^2r^6}+\frac{16\alpha Q^4}{\beta_m^2r^8}-\frac{15\alpha Q^4}{\beta_m^2r^9}+\Big(\frac{1}{72}+14\alpha\beta_m^2\Big)\frac{Q^6}{\beta_m^4r^{10}}+\mathcal{O}(r^{-12})\,.\label{r232}
\end{align}
\end{widetext}
One can compare Eqs.~\eqref{r232} with the approximated metric functions of the Einstein-Born-Infeld black hole \eqref{329} to see how the additional $R^2$ term in the action modifies the series of the metric functions.

\section{The EiBI gravity coupled with Maxwell fields}\label{sectIV}
In the previous section, we consider charged black holes within the Palatini $f(R)$ gravity coupled with Born-Infeld NED. The deviations of the solutions from the standard charged black holes in GR are essentially from both gravitational part ($f(R)$ modifications) and matter part (Born-Infeld NED). If we focus on the Einstein-Born-Infeld black hole whose metric functions are given in Eqs.~\eqref{EBIpsif}, the large curvature modification is purely from the matter sector. However, it is possible to consider a modified gravity model with a Born-Infeld correction from the gravity sector. Within the Palatini variational principle, it is intuitively the EiBI gravity.

In this section, we will review the exact solutions of the charged black holes within the EiBI gravity coupled with Maxwell fields. The solutions contain a Born-Infeld correction from the gravitational part of the action. It is interesting to investigate how the Born-Infeld correction from matter sector and from gravitational sector leads to different features of the QNM frequencies. We should emphasize that the exact expression of the charged black hole in the EiBI gravity has been derived in Refs.~\cite{Wei:2014dka,Sotani:2014lua}. In this section we will briefly review the derivation and recast the expressions in such a way that they can be better applied to calculate the QNM frequencies.

\subsection{Equations of motion}
The action of the EiBI model is \cite{Banados:2010ix}
\begin{equation}
\mathcal{S}_{2}=\frac{\epsilon\beta_g^2}{8\pi}\int d^4x\Big(\sqrt{\Big|g_{\mu\nu}+\frac{R_{(\mu\nu)}}{\epsilon\beta_g^2}\Big|}-\lambda\sqrt{-g}\Big)+\mathcal{S}_m.
\end{equation}
In the above action, $\epsilon=\pm1$ indicates that the Born-Infeld coupling constant can be either positive or negative. The dimensionless constant $\lambda$ is related to an effective cosmological constant by $\Lambda=\epsilon\beta_g^2(\lambda-1)$. Note that only the symmetric part of the Ricci tensor $R_{(\mu\nu)}(\Gamma)$ is considered in the action.

The field equation obtained by varying the action with respect to $g$ is
\begin{equation}
\sqrt{-q}q^{\mu\nu}-\lambda\sqrt{-g}g^{\mu\nu}=-\frac{8\pi}{\epsilon\beta_g^2}\sqrt{-g}T^{\mu\nu},
\label{eqeibi}
\end{equation}
where $T^{\mu\nu}$ is the energy momentum tensor and $q^{\mu\nu}$ is the inverse of the auxiliary metric $q_{\mu\nu}$, which is defined by $q_{\mu\nu}=g_{\mu\nu}+(\epsilon R_{(\mu\nu)}/\beta_g^2)$. The purpose of introducing the auxiliary metric $q_{\mu\nu}$ is due to the variation with respect to the affine connection. It can be shown that the affine connection can be chosen to be the Levi-Civita connection of the auxiliary metric $q_{\mu\nu}$. 

If the theory is coupled with the Maxwell electromagnetic fields, the energy momentum tensor can be derived by simply inserting $\phi=X$ in Eq.~\eqref{nedTmunu}:
\begin{equation}
T_{\mu\nu}=\frac{1}{4\pi}\Big(F_{\mu\sigma}{F_{\nu}}^\sigma-\frac{1}{4}F_{\alpha\beta}F^{\alpha\beta} g_{\mu\nu}\Big).
\end{equation}
Since we only consider the purely radial electric field, Eq.~\eqref{21} is valid and the energy momentum tensor can be written as
\begin{equation}
{T_{\mu}}^{\nu}=\frac{Q^2}{8\pi r^4}
\begin{bmatrix}
    -\hat{I}_{2\times 2}       & \hat{0} \\
    \hat{0}      &\hat{I}_{2\times 2}
    \end{bmatrix},
    \label{contractT2}
\end{equation}
where $\hat{I}_{2\times2}$ is a two-by-two identity matrix. Then we define a matrix $\hat{\Omega}$ as a map between the two metrics: $\hat{\Omega}\equiv\hat{g}^{-1}\hat{q}$ such that $\hat{q}=\hat{g}\hat{\Omega}$. According to the definition of $q_{\mu\nu}$, one can define the auxiliary curvature as follows
\begin{equation}
{R_{\mu}}^{\nu}(q)\equiv q^{\nu\alpha}R_{(\mu\alpha)}(\Gamma)=\epsilon\beta_g^2(\hat{I}-\hat{\Omega}^{-1}).
\label{Rqq}
\end{equation}
The matrix $\hat{\Omega}$ can be obtained via Eqs.~\eqref{eqeibi} and \eqref{contractT2}:
\begin{equation}
\hat{\Omega}=
\begin{bmatrix}
    \sigma_{-}\hat{I}_{2\times 2}       & \hat{0} \\
    \hat{0}      &\sigma_{+}\hat{I}_{2\times 2}
    \end{bmatrix},
    \label{omega}
\end{equation}
where 
\begin{equation}
\sigma_{\pm}\equiv\lambda\pm\frac{Q^2}{\epsilon\beta_g^2r^4}.
\end{equation}

Considering again the metric ansatz:
\begin{equation}
ds^2_g=-\psi^2(r)\bar{f}(r)dt^2+\frac{1}{\bar{f}(r)}dr^2+r^2d\Omega^2,
\label{gmetricreeibi}
\end{equation}
and the line element of $q_{\mu\nu}$
\begin{equation}
ds^2_q=-G^2(x)F(x)dt^2+\frac{1}{F(x)}dx^2+x^2d\Omega^2,
\label{qmetric}
\end{equation}
we have following identities:
\begin{equation}
G^2(x)F(x)=\psi^2\bar{f}\sigma_{-}\,,\qquad\Big(\frac{dx}{dr}\Big)^2=\sigma_{-}\frac{F(x)}{\bar{f}(r)}\,,
\end{equation}
and 
\begin{equation}
x^2=r^2\sigma_{+}.
\end{equation}
From the last two identities, we have
\begin{equation}
\frac{dx}{dr}=\frac{\sigma_{-}}{\sqrt{\sigma_{+}}}\,,\qquad F=\bar{f}\frac{\sigma_{-}}{\sigma_{+}}\,.
\label{412}
\end{equation}

By calculating the non-vanishing components of Eq.~\eqref{Rqq}, we can obtain 
\begin{equation}
G=\textrm{constant},
\end{equation}
which we choose $G=\sqrt{\lambda}$ after a time rescaling. In this regard, the metric function $\psi$ can be obtained
\begin{equation}
\psi^2=\frac{r^4}{r^4+\frac{\epsilon Q^2}{\beta_g^2\lambda}}.
\label{psi2eibi}
\end{equation}
The other metric function can be derived through the definition of the mass function $M$
\begin{equation}
F=1-\frac{2M(x)}{x},\label{415}
\end{equation}
where the mass function satisfies
\begin{equation}
\frac{dM(x)}{dx}=\frac{x^2\epsilon\beta_g^2}{2}\Big(1-\frac{1}{\sigma_{+}}\Big).
\end{equation}
After an integration and assuming $\lambda=1$ (a vanishing cosmological constant), we have
\begin{align}
M(r)=\frac{r_s}{2}-\frac{\epsilon\beta_g^2}{2}\int_r^{\infty}r^2(\sigma_{+}-1)\frac{\sigma_{-}}{\sqrt{\sigma_{+}}}dr.
\end{align}
Then we apply the following rescalings
\begin{equation}
\frac{Q}{r_s}\rightarrow Q\qquad \beta_gr_s\rightarrow \beta_g\qquad\frac{r}{r_s}\rightarrow r\,,
\end{equation}
and the mass function becomes
\begin{align}
M(r)=M_0\Big[1-\epsilon\beta_g^2\int_r^{\infty}r^2(\sigma_{+}-1)\frac{\sigma_{-}}{\sqrt{\sigma_{+}}}dr\Big].\label{masseibi}
\end{align}
To proceed, we define a dimensionless radius $z=r/r_g$ where $r_g^4\equiv Q^2/\beta_g^2$, as what we have done in the previous section.

\subsection{Metric functions}
\subsubsection{$\epsilon=+1$}
If $\epsilon=+1$, the metric function $\psi^2$ in Eq.~\eqref{psi2eibi} is
\begin{equation}
\psi^2=\frac{1}{1+\frac{1}{z^4}}=\frac{r^4}{r^4+r_g^4}.
\end{equation}
On the other hand, the metric function $\bar{f}$ can be obtained via Eqs.~\eqref{412}, \eqref{415} and \eqref{masseibi} \cite{integrals,integrals2}
\begin{widetext}
\begin{align}
\bar{f}&=\frac{\sigma_{+}}{\sigma_{-}}F\nonumber\\
&=\frac{r^4+r_g^4}{r^4-r_g^4}\Big[1-\frac{r}{\sqrt{r^4+r_g^4}}\Big(1-\frac{2r_g^3\beta_g^2}{3}F\Big(\cos^{-1}{\frac{r^2-r_g^2}{r^2+r_g^2}\,,\frac{1}{\sqrt{2}}}\Big)\Big)-\frac{r_g^4\beta_g^2}{3r^2}\Big]\nonumber\\
&=\frac{r^4+r_g^4}{r^4-r_g^4}\Big[1-\frac{r}{\sqrt{r^4+r_g^4}}\Big(1-\frac{4r_g^4\beta_g^2}{3r}F\Big(\frac{1}{4},\frac{1}{2},\frac{5}{4},-\frac{r_g^4}{r^4}\Big)\Big)-\frac{r_g^4\beta_g^2}{3r^2}\Big].\label{EiBIplus}
\end{align}
\end{widetext}

Furthermore, the asymptotic expansions of the metric functions at $r\gg r_g$ read

\begin{widetext}
\begin{align}
\psi^2\bar{f}&=1-\frac{1}{r}+\frac{Q^2}{r^2}+\frac{Q^2}{\beta_g^2r^4}-\frac{Q^2}{2\beta_g^2r^5}+\frac{Q^4}{5\beta_g^2r^6}+\frac{Q^4}{\beta_g^4r^8}-\frac{7Q^4}{8\beta_g^4r^9}+\frac{37Q^6}{45\beta_g^4r^{10}}+\mathcal{O}(r^{-12}),\nonumber\\
\bar{f}&=1-\frac{1}{r}+\frac{Q^2}{r^2}+\frac{2Q^2}{\beta_g^2r^4}-\frac{3Q^2}{2\beta_g^2r^5}+\frac{6Q^4}{5\beta_g^2r^6}+\frac{2Q^4}{\beta_g^4r^8}-\frac{11Q^4}{8\beta_g^4r^9}+\frac{46Q^6}{45\beta_g^4r^{10}}+\mathcal{O}(r^{-12}).\label{423}
\end{align}
\end{widetext}
The solution is again approximately the standard RN black hole, as it should be when $r\gg r_g$.

\subsubsection{$\epsilon=-1$}
If $\epsilon=-1$, the metric function $\psi^2$ in Eq.~\eqref{psi2eibi} is
\begin{equation}
\psi^2=\frac{1}{1-\frac{1}{z^4}}=\frac{r^4}{r^4-r_g^4}.
\end{equation}
On the other hand, the metric function $\bar{f}$ can be obtained via Eqs.~\eqref{412}, \eqref{415} and \eqref{masseibi} \cite{integrals,integrals2}
\begin{widetext}
\begin{equation}
\bar{f}=\frac{r^4-r_g^4}{r^4+r_g^4}\Big[1-\frac{r}{\sqrt{r^4-r_g^4}}\Big(1-\frac{r_g^3\beta_g^2}{3}B\Big(\frac{1}{4},\frac{1}{2}\Big)+\frac{2\sqrt{2}r_g^3\beta_g^2}{3}F\Big(\cos^{-1}\frac{r_g}{r},\frac{1}{\sqrt{2}}\Big)\Big)-\frac{r_g^4\beta_g^2}{3r^2}\Big],\label{EiBIminus}
\end{equation}
\end{widetext}
where $B(..,..)$ is the Beta function. In this case, it can be easily seen that the radius $r$ has a minimum value at $r=r_g$. 

The asymptotic expressions of the metric functions at $r\gg r_g$ are
\begin{widetext}\label{427}
\begin{align}
\psi^2\bar{f}&=1-\frac{1}{r}+\frac{Q^2}{r^2}-\frac{Q^2}{\beta_g^2r^4}+\frac{Q^2}{2\beta_g^2r^5}-\frac{Q^4}{5\beta_g^2r^6}+\frac{Q^4}{\beta_g^4r^8}-\frac{7Q^4}{8\beta_g^4r^9}+\frac{37Q^6}{45\beta_g^4r^{10}}+\mathcal{O}(r^{-12}),\nonumber\\
\bar{f}&=1-\frac{1}{r}+\frac{Q^2}{r^2}-\frac{2Q^2}{\beta_g^2r^4}+\frac{3Q^2}{2\beta_g^2r^5}-\frac{6Q^4}{5\beta_g^2r^6}+\frac{2Q^4}{\beta_g^4r^8}-\frac{11Q^4}{8\beta_g^4r^9}+\frac{46Q^6}{45\beta_g^4r^{10}}+\mathcal{O}(r^{-12}).\label{427}
\end{align}
\end{widetext}
It is again apparent to see that the solution reduces to the standard RN black hole, as expected when $r\gg r_g$.

\begin{figure}
\centering
\graphicspath{{fig/}}
\includegraphics[scale=0.47]{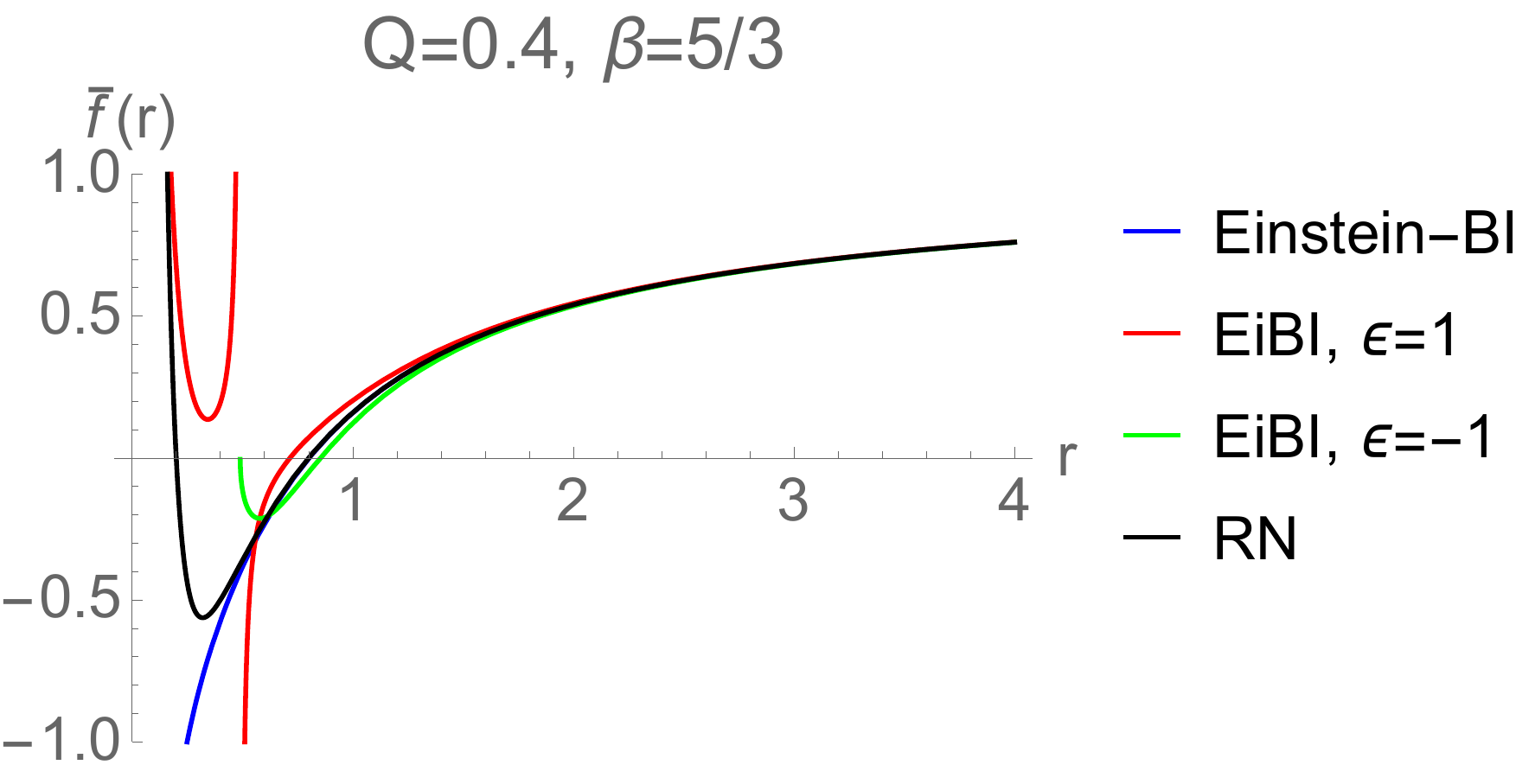}
\includegraphics[scale=0.44]{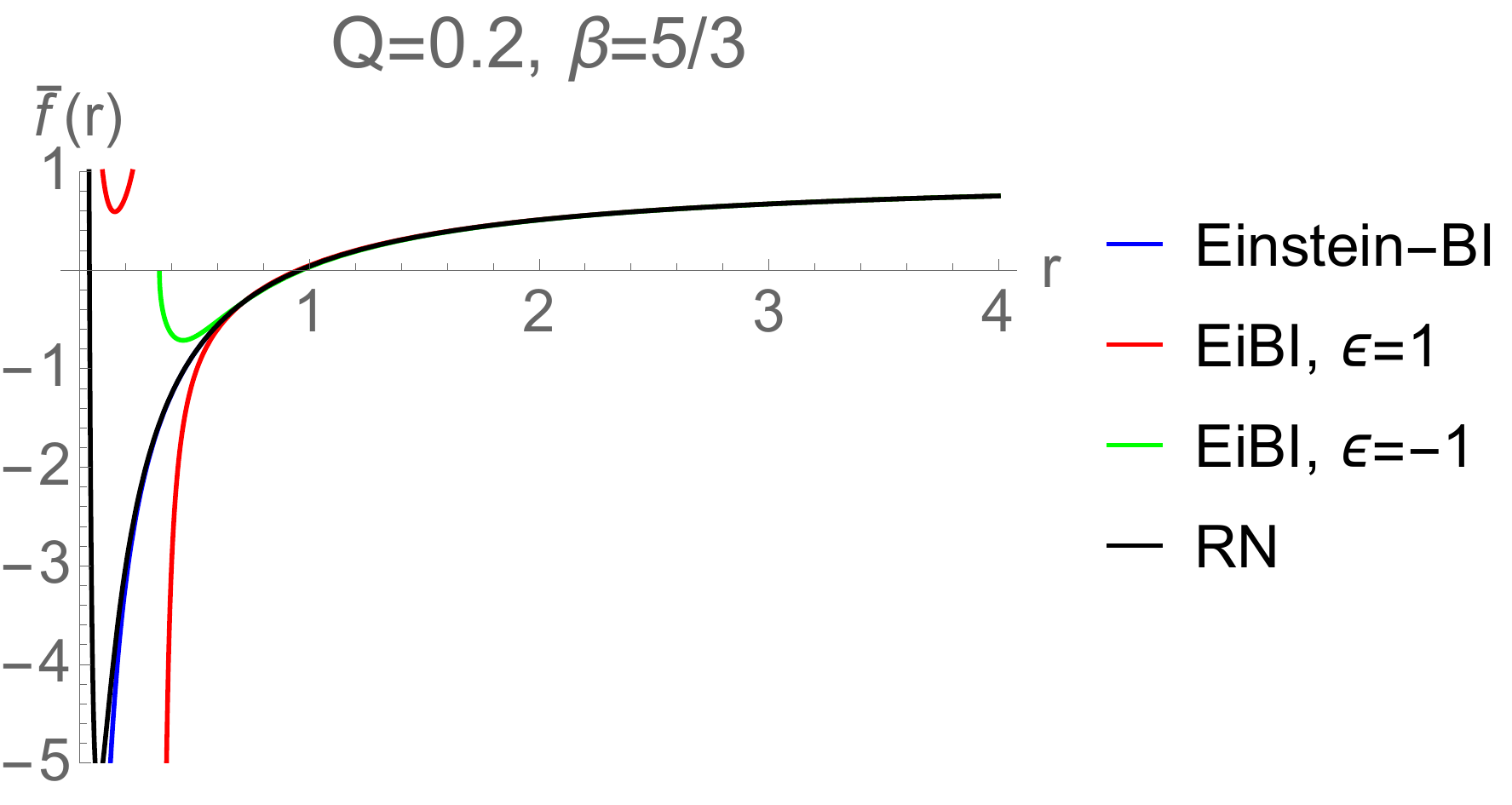}
\caption{The exact metric function $\bar{f}(r)$ of different types of black holes. Note that only the parameter space where the black hole horizon exists is considered.} 
\label{fig}
\end{figure}

In FIG.~\ref{fig}, we exhibit the metric functions of the black hole solutions considered so far. It can be seen that even though all solutions reduce to RN black holes at large radius, the metric functions would behave significantly differently inside the event horizon. For example, the interior solution of the EiBI black hole with $\epsilon=-1$ has been shown to be more regular than the RN black hole and is geodesically complete \cite{Olmo:2013gqa,Olmo:2015bya,Olmo:2015dba}. We would like to stress that due to the Born-Infeld corrections, there in principle are some parameter spaces where no black hole exists \cite{Wei:2014dka,Sotani:2014lua}. In this work, we only consider the cases where the black holes exist and calculate their QNM frequencies.  

\section{QNM frequencies: The 6th order WKB method}\label{sectV}
Technically, there are several methods to calculate the QNM frequencies, ranging from purely numerical approaches \cite{Leaver:1986gd,Jansen:2017oag} to some semi-analytic methods (see Refs~\cite{Nollert:1999ji,Berti:2009kk,Konoplya:2011qq,Berti:2015itd} and references therein). In this work, we will use one of the latter based on the WKB approximation. This method was firstly formulated in a seminal paper \cite{Schutz:1985zz} and one of the advantages of using the WKB technique is that the QNM frequencies can be calculated with one simple formula, as long as an effective potential associated with the exterior side of the black hole is given. In Refs.~\cite{Iyer:1986np,Konoplya:2003ii}, the first order WKB method was further extended to 3rd order and 6th order WKB formula, respectively. Recently, the extension of the WKB method to the 13th order has also been formulated by using the Pad\'e transforms \cite{Matyjasek:2017psv}. This method has already been applied in plenty of works and has also been confirmed to be accurate at least when the multipole number $l$ is larger than the overtone $n$ \cite{Berti:2009kk}. For the astrophysical black holes, this is compatible in the sense that the fundamental mode $n=0$ has a longest decay time and it dominates the late time signal of the ringdown stage. Therefore, the analysis in this work will mostly focus on the fundamental mode and on those QNMs whose multipole numbers satisfy $l>n$.  

Let us briefly elucidate this semi-analytic method which is based on the WKB technique. The WKB technique is inspired by the fact that the perturbation equation \eqref{perturbationeq} (not only for perturbations driven by scalar fields, but also those driven by other test fields with different spin) resembles the Schr\"{o}dinger wave equation in quantum mechanics. The effective potential $V_{\textrm{eff}}(r_*)\equiv-\omega^2+V(r(r_*))$, in most cases (including ours), has finite value when $r_*\rightarrow\infty$ (spatial infinity) and $r_*\rightarrow-\infty$ (at the event horizon). In our case, the effective potential approaches $-\omega^2$ at $r_*=\pm\infty$. Furthermore, $V_{\textrm{eff}}(r_*)$ has a maximum value at some finite $r_*$. Therefore, the problem can be regarded as a quantum scattering process through a potential barrier, as long as some proper boundary conditions are taken into account. The appropriate boundary conditions associated with the black hole geometry can be comprehended from the perspective of an observer at spatial infinity and at the event horizon. At spatial infinity, it is required that only outgoing waves moving away from the black hole exist. At the event horizon, on the other hand, there can only be ingoing waves moving toward the black hole for a physically acceptable solution. 

To encompass these boundary conditions, one can consider a quantum scattering process without incident waves, while the reflected and the transmitted waves have comparable amounts of amplitudes. Following Ref.~\cite{Schutz:1985zz}, it is required that the maximum value of the effective potential $V_{\textrm{eff}}(r_*)$ is slightly larger than zero and there are two classical turning points at the vicinity of the peak. At the regions far away from the turning points ($r_*\rightarrow\pm\infty$), the solutions are calculated using the WKB approximation up to a desired order and the aforementioned boundary conditions should be considered. Near the peak, the effective potential is expanded into a Taylor series up to a corresponding order, and one can solve the differential equation by using the series expansion of the potential. Then, by matching the power series of the solution near the peak with the solutions obtained with the WKB approximation simultaneously at the turning points, the numerical values of the QNM frequencies $\omega$ can be derived according to the matching conditions.

In the 6th order WKB method, in general, the WKB formula for calculating QNM frequencies is \cite{Konoplya:2003ii}
\begin{equation}
\frac{i(\omega^2-V_m)}{\sqrt{-2V''_m}}-\sum_{i=2}^6\Lambda_i=n+\frac{1}{2},
\end{equation}
where $n$ is the overtone number, and the index $m$ denotes the quantity evaluated at the peak of the potential. $V''_m$ is the second order derivative of the potential with respect to $r_*$, calculated at the peak. $\Lambda_i$ are constant coefficients from higher order WKB corrections and they depend on the value and derivatives (up to 12th order) of the potential at the peak.{\footnote{The explicit expressions of $\Lambda_i$ are given in Refs.~\cite{Iyer:1986np,Konoplya:2003ii}}} 

\subsection{Fundamental QNM frequencies}

Before calculating the QNM frequencies of different charged black holes, we want to highlight that in the WKB technique, only the spacetime outside the horizon contributes to the evaluation of the QNM frequencies. In FIG.~\ref{fig}, it can be seen that the deviations of the metric functions considered here from the RN charged black hole are very tiny outsider the event horizon, even though the metric functions in different models would behave drastically differently insider the horizon. Therefore, instead of using the full expressions of the metric functions Eqs.~\eqref{EBIpsif}, \eqref{EiBIplus}, and \eqref{EiBIminus}, we will use their approximated forms \eqref{329}, \eqref{423}, and \eqref{427} to calculate the QNM frequencies. This can prevent us from struggling in higher order derivatives of the special functions and shorten the calculation time. Furthermore, these approximated expressions are accurate enough not only for large radius, but also for small $Q/\beta$, which is believed to be observationally acceptable for astrophysical black holes. 

\begin{figure}
\centering
\graphicspath{{fig/}}
\includegraphics[scale=0.35]{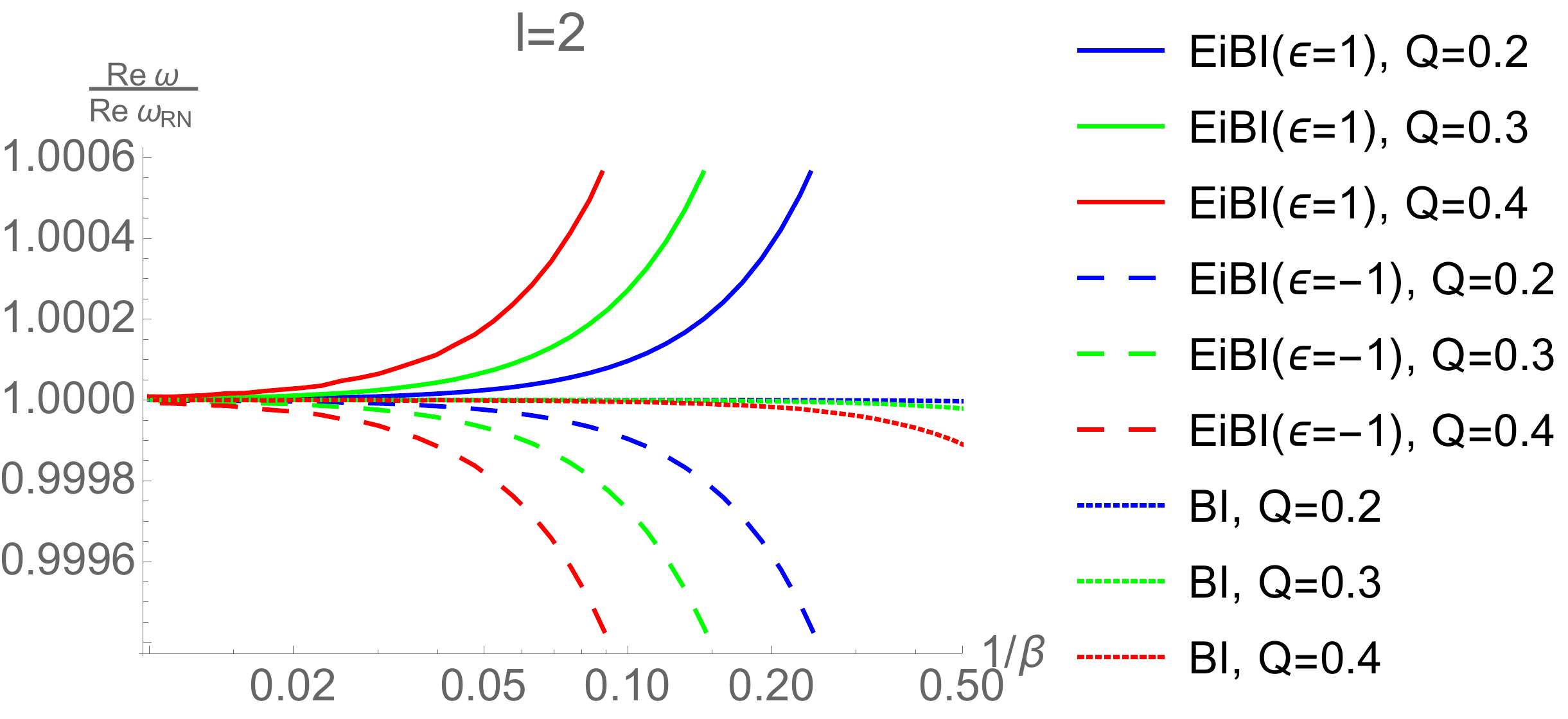}
\includegraphics[scale=0.35]{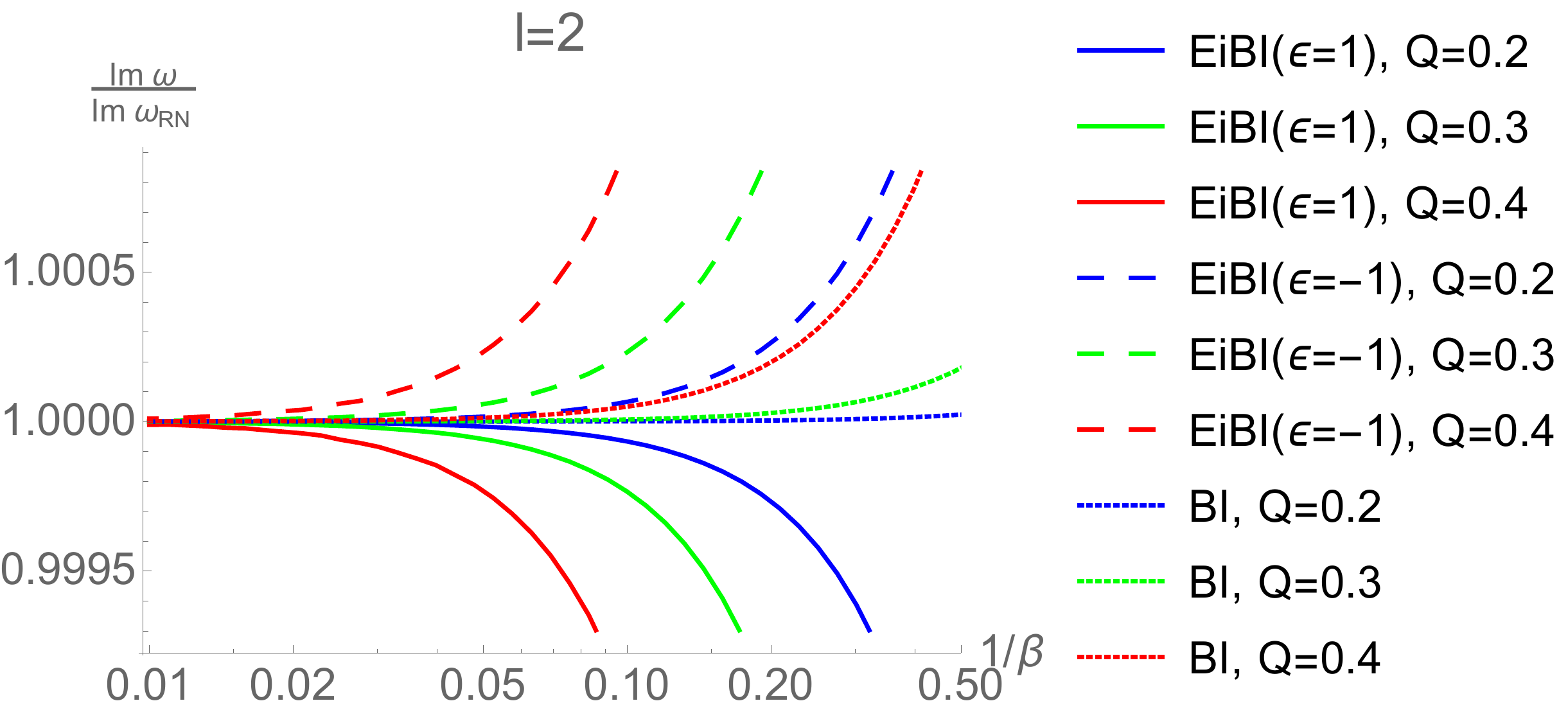}
\caption{The QNM frequency of Einstein-BI black holes (dotted) and EiBI charged black holes with a positive (solid) and a negative (dashed) coupling constant $\epsilon$. The frequencies are shown as a function of $1/\beta$ ($\beta_m$ for Einstein-BI black holes and $\beta_g$ for EiBI black holes). We consider the fundamental mode $n=0$ and $l=2$. The curves in different colors represent different values of charge.} 
\label{fig1}
\end{figure}

\begin{figure}
\centering
\graphicspath{{fig/}}
\includegraphics[scale=0.35]{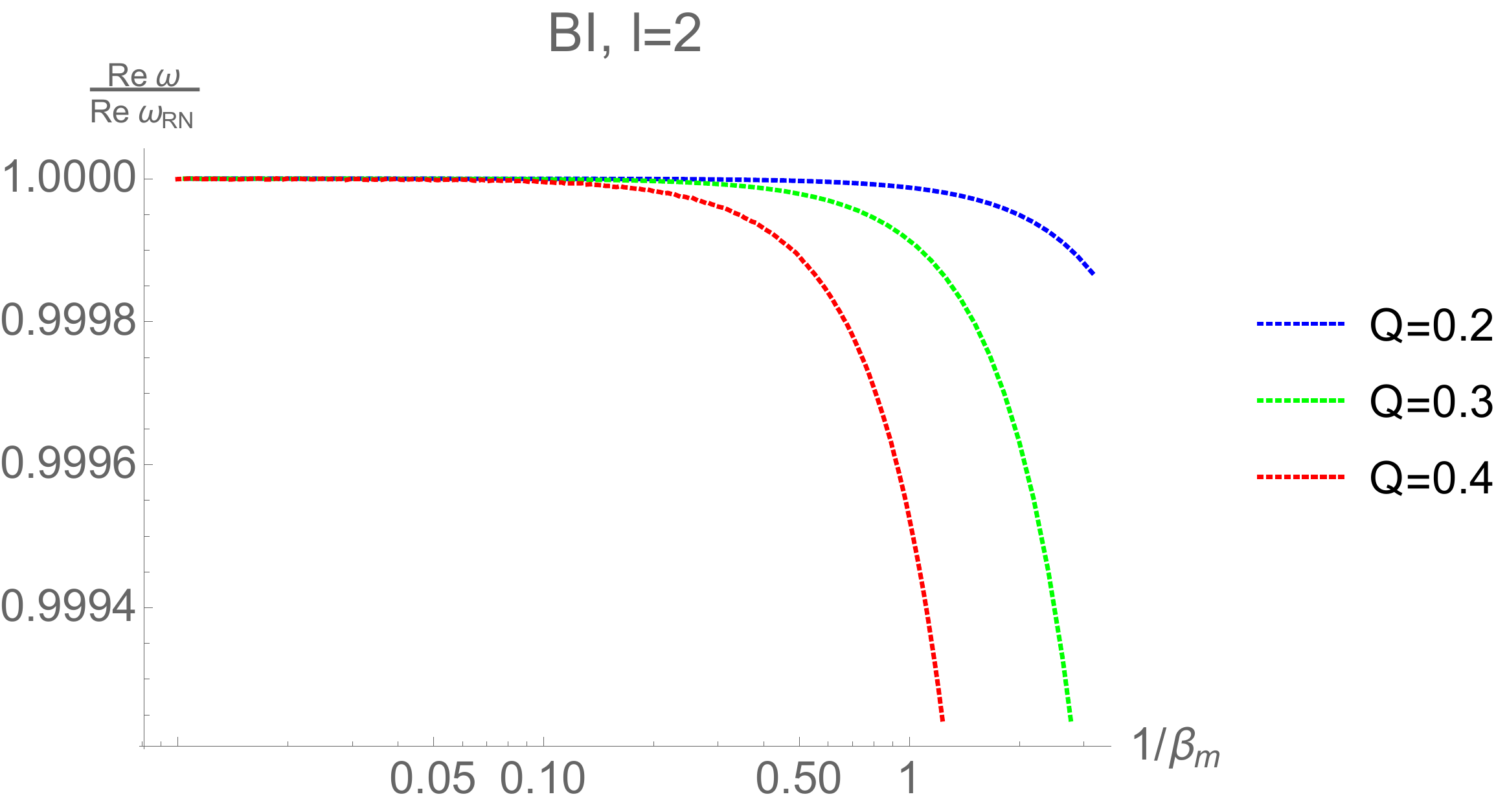}
\includegraphics[scale=0.35]{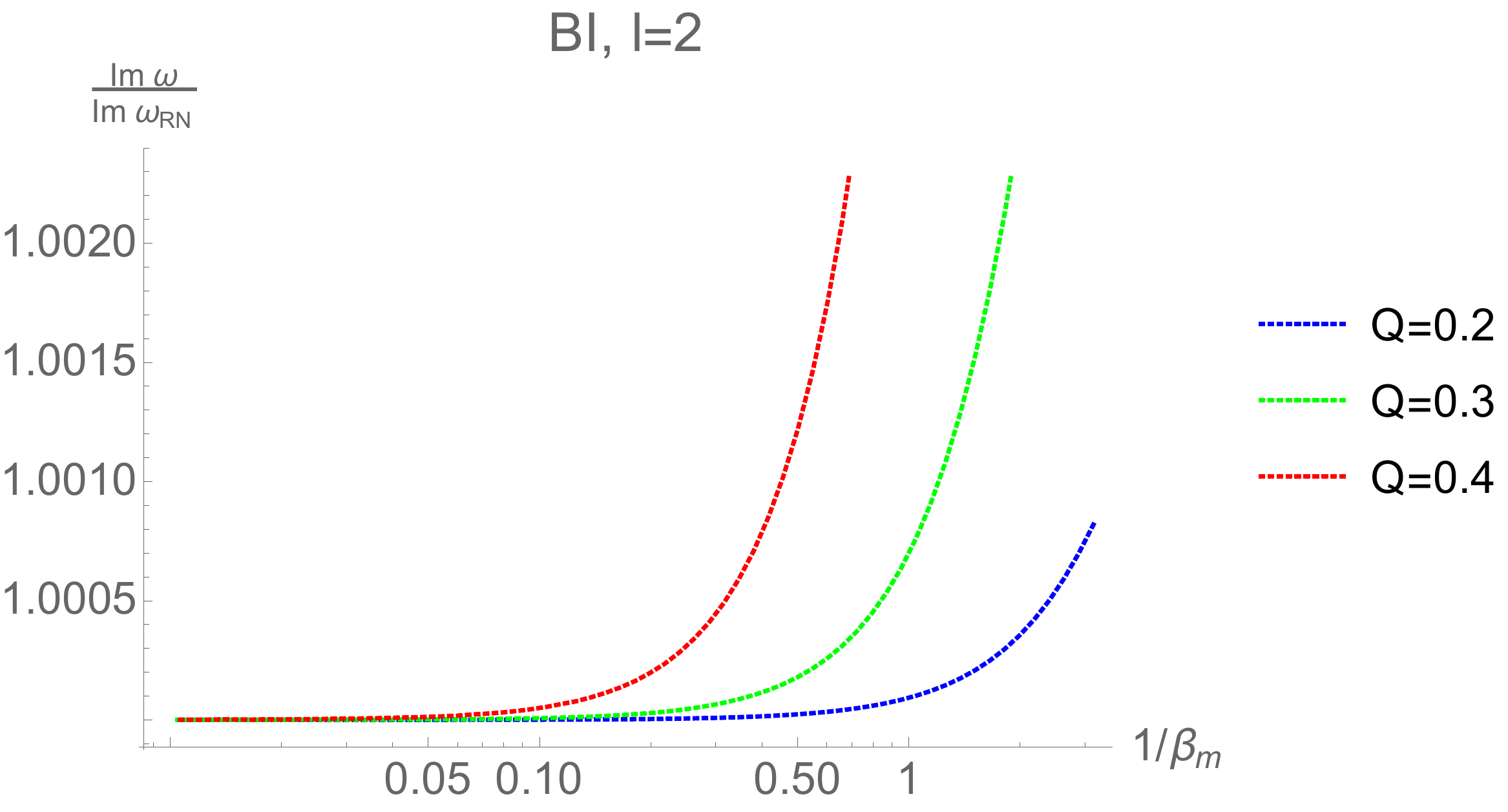}
\caption{The QNM frequency of Einstein-BI black holes as a function $1/\beta_m$. We consider the fundamental mode $n=0$ and $l=2$. The curves in different colors represent different values of charge.} 
\label{fig2}
\end{figure}

In FIG.~\ref{fig1}, we exhibit the real part (upper) and the imaginary part (lower) of the QNM frequency of Einstein-Born-Infeld black holes (dotted) and EiBI charged black holes with positive (solid) and negative (dashed) coupling constant in terms of the Born-Infeld corrections $1/\beta$ ($\beta_m$ for Einstein-BI black holes and $\beta_g$ for EiBI black holes). We consider the multiple $l=2$ and the fundamental mode $n=0$. To highlight the deviations due to the Born-Infeld corrections, we present the QNM frequency ratio of the modified black hole and the standard RN charged black hole. Some qualitative conclusions are put forward as follows:
\begin{itemize}
\item A smaller charge corresponds to smaller deviations from $\omega_{\textrm{RN}}$, as expected.
\item For EiBI charged black holes with $\epsilon>0$ ($\epsilon<0$), the real part of $\omega$ increases (decreases) with $1/\beta_g$. The decay time, which is proportional to $1/|\textrm{Im }\omega|$, also increases (decreases) with $1/\beta_g$.

\item For Einstein BI black holes, the real part of $\omega$ and the decay time decrease with $1/\beta_m$. A much clearer tendency can be understood in FIG.~\ref{fig2}.{\footnote{Our results for the imaginary part of $\omega$ are qualitatively consistent with Ref.~\cite{Fernando:2005bc}. However, because in \cite{Fernando:2005bc} the authors chose a much smaller value of $\beta_m$, the tendency of real part of $\omega$ could be different.}}

\item The deviations of the Einstein BI black hole frequency from the GR counterpart are less sensitive to the change in $\beta$ than those of the EiBI charged black hole. This may be understood from the fact that the lowest order containing $\beta_m$ corrections in Eq.~\eqref{329} is $1/r^6$ term. But in the EiBI charged black hole, the corrections appear in $1/r^4$ term. 
\end{itemize}

\begin{figure}
\centering
\graphicspath{{fig/}}
\includegraphics[scale=0.36]{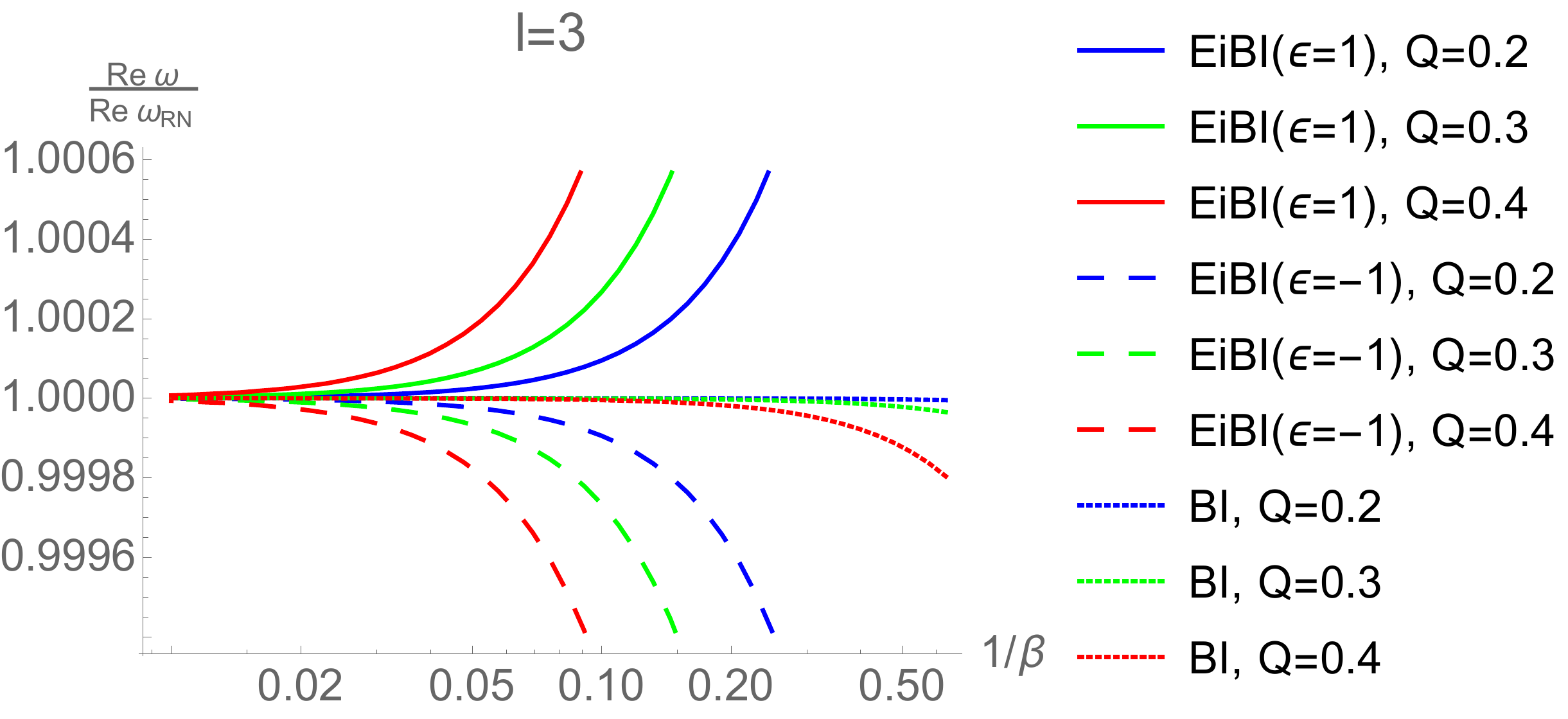}
\includegraphics[scale=0.36]{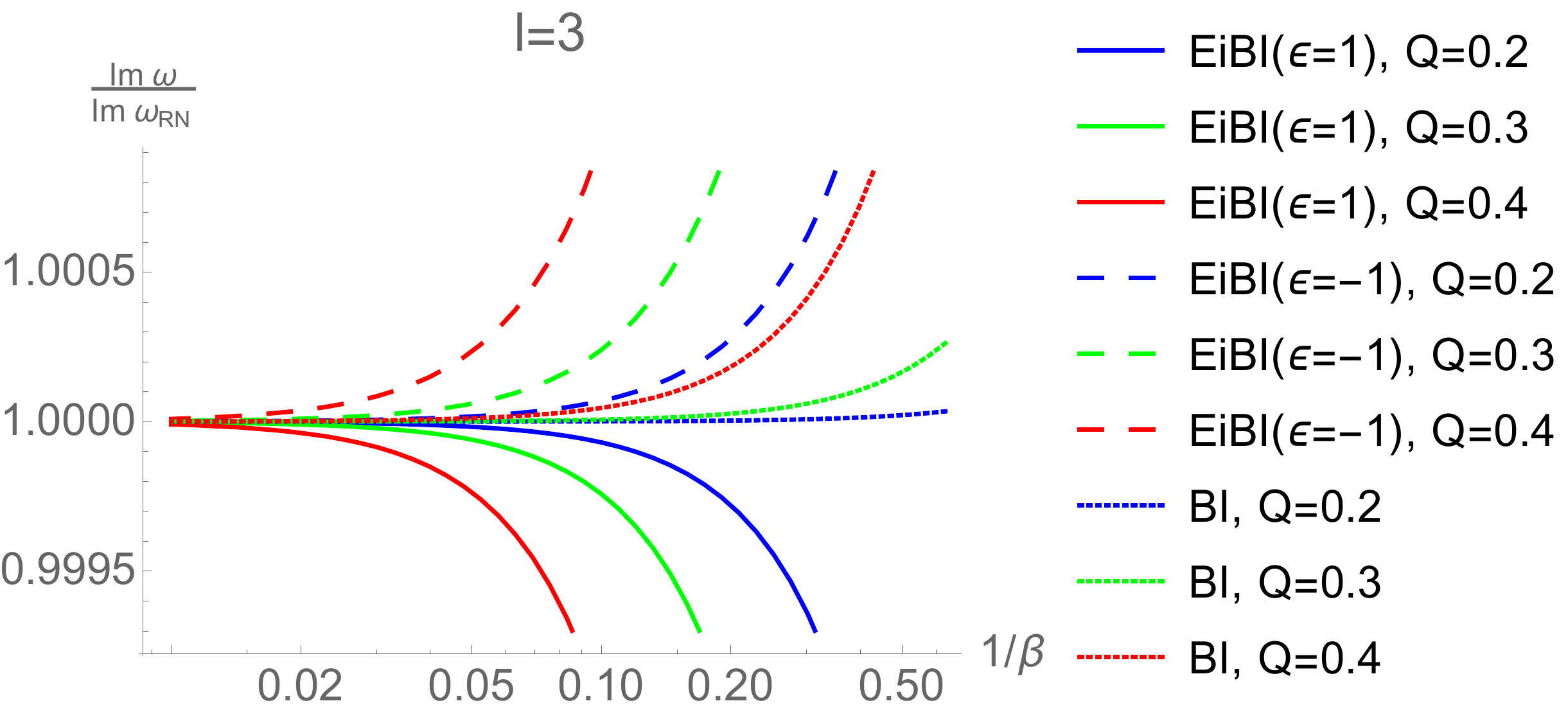}
\caption{The QNM frequency of Einstein-BI black holes (dotted) and EiBI charged black holes with a positive (solid) and a negative (dashed) coupling constant $\epsilon$. The frequencies are shown as a function of $1/\beta$ ($\beta_m$ for Einstein-BI black holes and $\beta_g$ for EiBI black holes). We consider the fundamental mode $n=0$ and $l=3$. The curves in different colors represent different values of charge.} 
\label{fig3}
\end{figure}

In addition to the multiple $l=2$, we also present the QNM frequency of these charged black holes with multiple $l=3$ in FIG.~\ref{fig3}. The above qualitative results are still valid.

\subsection{Eikonal QNMs}
By nature the WKB technique is more accurate when $l>n$ as mentioned previously. Therefore, it is straightforward to consider the QNM frequency in the eikonal limit ($l\rightarrow\infty$). According to Ref.~\cite{Cardoso:2008bp}, the QNM frequency in the eikonal limit can be expressed as
\begin{equation}
\omega\approx\Omega_cl-i(n+1/2)|\lambda_c|,
\label{eikonal}
\end{equation}
where
\begin{align}
\Omega_c&=\frac{\psi(r_c)\sqrt{\bar{f}(r_c)}}{r_c}\,,\\
\lambda_c&=\frac{1}{\sqrt{2}}\sqrt{-\frac{r_c^2}{\psi^2(r_c)\bar{f}(r_c)}\Big[\frac{d^2}{dr_*^2}\Big(\frac{\psi^2\bar{f}}{r^2}\Big)\Big]_{r=r_c}}\,,
\end{align}
and $r_c$ is the radius of the null circular orbit around the black hole. In general, $r_c$ would satisfy \cite{Cardoso:2008bp}
\begin{equation}
2\psi^2(r_c)\bar{f}(r_c)=r_c\frac{d}{dr}(\psi^2(r)\bar{f}(r))\Big|_{r=r_c}.
\end{equation}

In fact, $\Omega_c$ can be interpreted as the angular velocity of a null circular orbit around the black hole. The parameter $\lambda_c$, on the other hand, is essentially the Lyapunov exponent quantifying the instability of the orbit. Note that a positive $\lambda_c$ corresponds to the instability of the dynamical system under consideration. The correspondence of the eikonal QNMs and the property of the null circular orbit around the black hole has been shown to be valid in many cases. It has been argued that this correspondence holds for any stationary, spherically symmetric, and asymptotically flat metric. Even though in some particular cases this correspondence would be violated \cite{Konoplya:2017wot,Toshmatov:2018tyo}, for a test scalar field in the charged black hole background considered in this paper, this correspondence is expected to be satisfied. 

Furthermore, it should be stressed that the derivation of the correspondence between eikonal QNMs and the parameters of the null circular orbit around the black hole, that is, Eq.~\eqref{eikonal}, is based on the validity of the WKB method proposed in \cite{Schutz:1985zz}. Therefore, the fulfillment of Eq.~\eqref{eikonal} requires the validity of the WKB technique. This means that in some cases where the effective potential might have local minima or more than one local maxima, the WKB technique breaks down and Eq.~\eqref{eikonal} is no longer satisfied.

\begin{figure}
\centering
\graphicspath{{fig/}}
\includegraphics[scale=0.41]{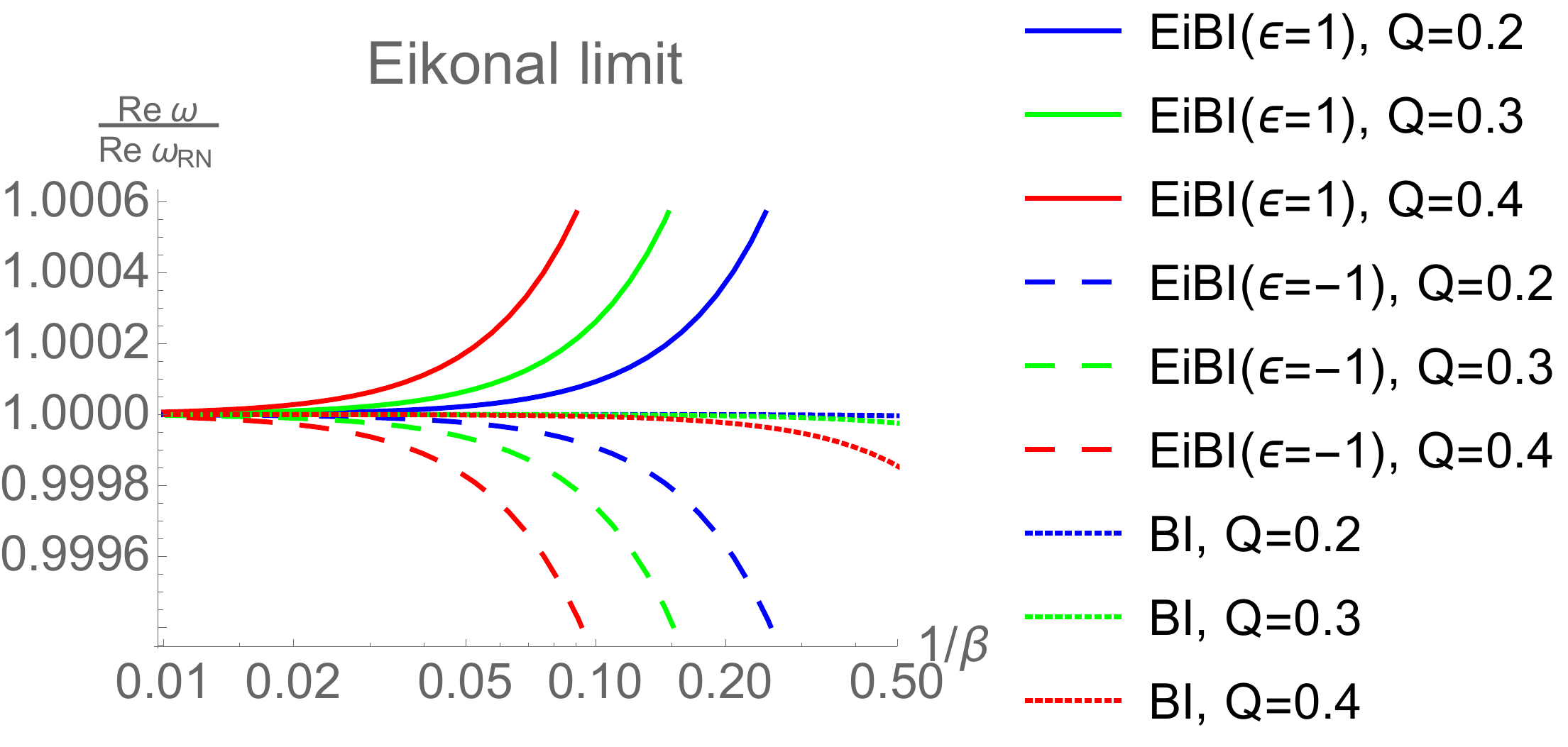}
\includegraphics[scale=0.41]{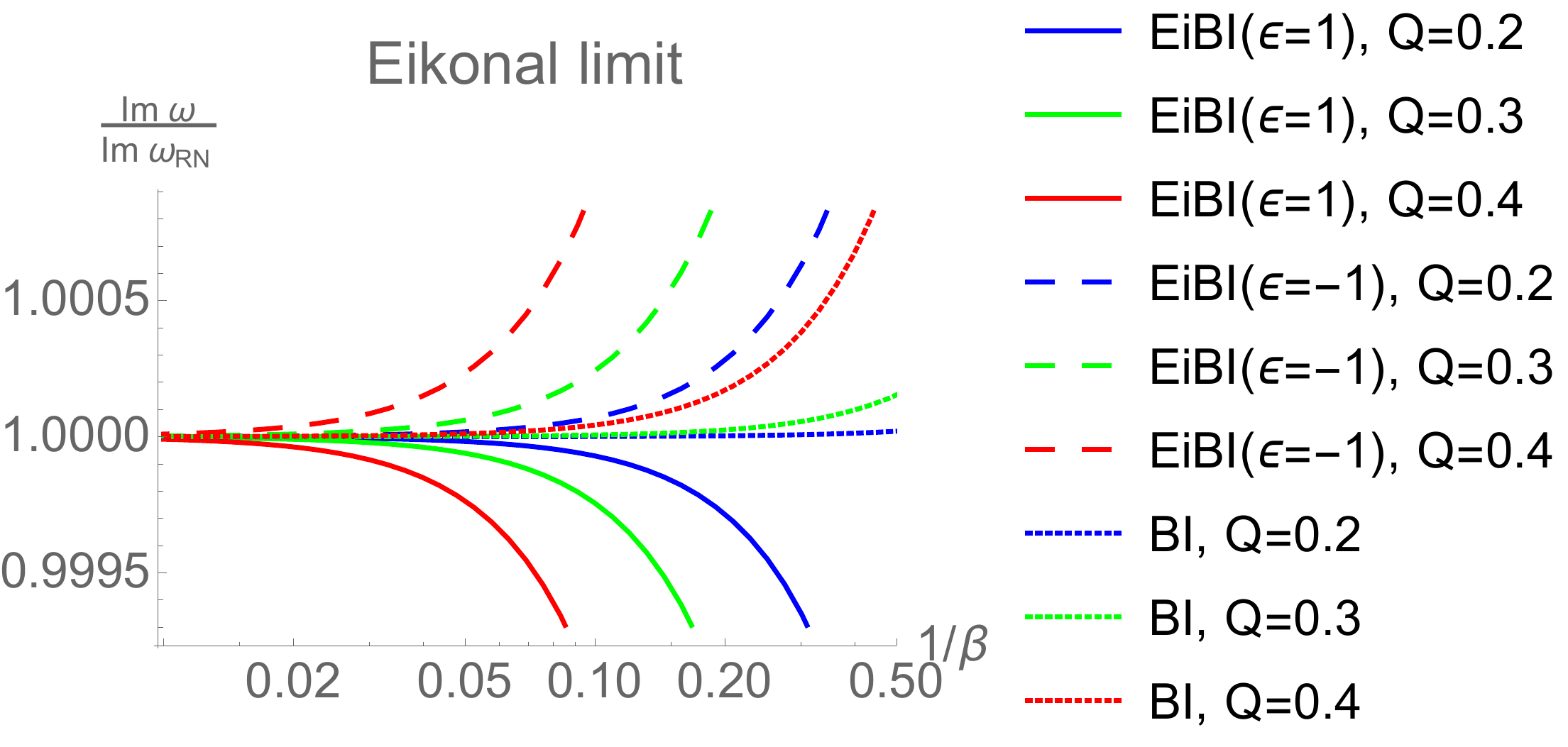}
\caption{The eikonal QNM frequency of Einstein-BI black holes (dotted) and EiBI charged black holes with a positive (solid) and a negative (dashed) coupling constant $\epsilon$. The frequencies are shown as a function of $1/\beta$ ($\beta_m$ for Einstein-BI black holes and $\beta_g$ for EiBI black holes). The curves in different colors represent different values of charge.} 
\label{eikonalfig}
\end{figure}

In FIG.~\ref{eikonalfig}, we present the real part (upper) and the imaginary part (lower) of the QNM frequencies within the eikonal limit. The frequencies are calculated by inserting the exact expressions of the metric functions Eqs.~\eqref{EBIpsif}, \eqref{EiBIplus}, and \eqref{EiBIminus} into the eikonal formula \eqref{eikonal}. It can be seen that the tendency of the deviations resulting from the Born-Infeld corrections are qualitatively similar to those with smaller multiple $l$ (see FIGs.~\ref{fig1} and \ref{fig3}). Note that here the frequencies are exhibited in the form of the ratio with the RN counterpart. Therefore, the results are independent of the multipole number $l$ and the overtone $n$.

\subsection{Palatini $R+\alpha R^2$ coupled with Born-Infeld NED}
Before closing this section, we would like to consider a more extended theory, i.e., a Palatini $R+\alpha R^2$ gravity coupled with Born-Infeld NED. Due to the complexity of the field equations, there is no exact expression of the metric functions. The metric functions can only be expressed in the integral form \eqref{integral}. 

\begin{figure}
\centering
\graphicspath{{fig/}}
\includegraphics[scale=0.5]{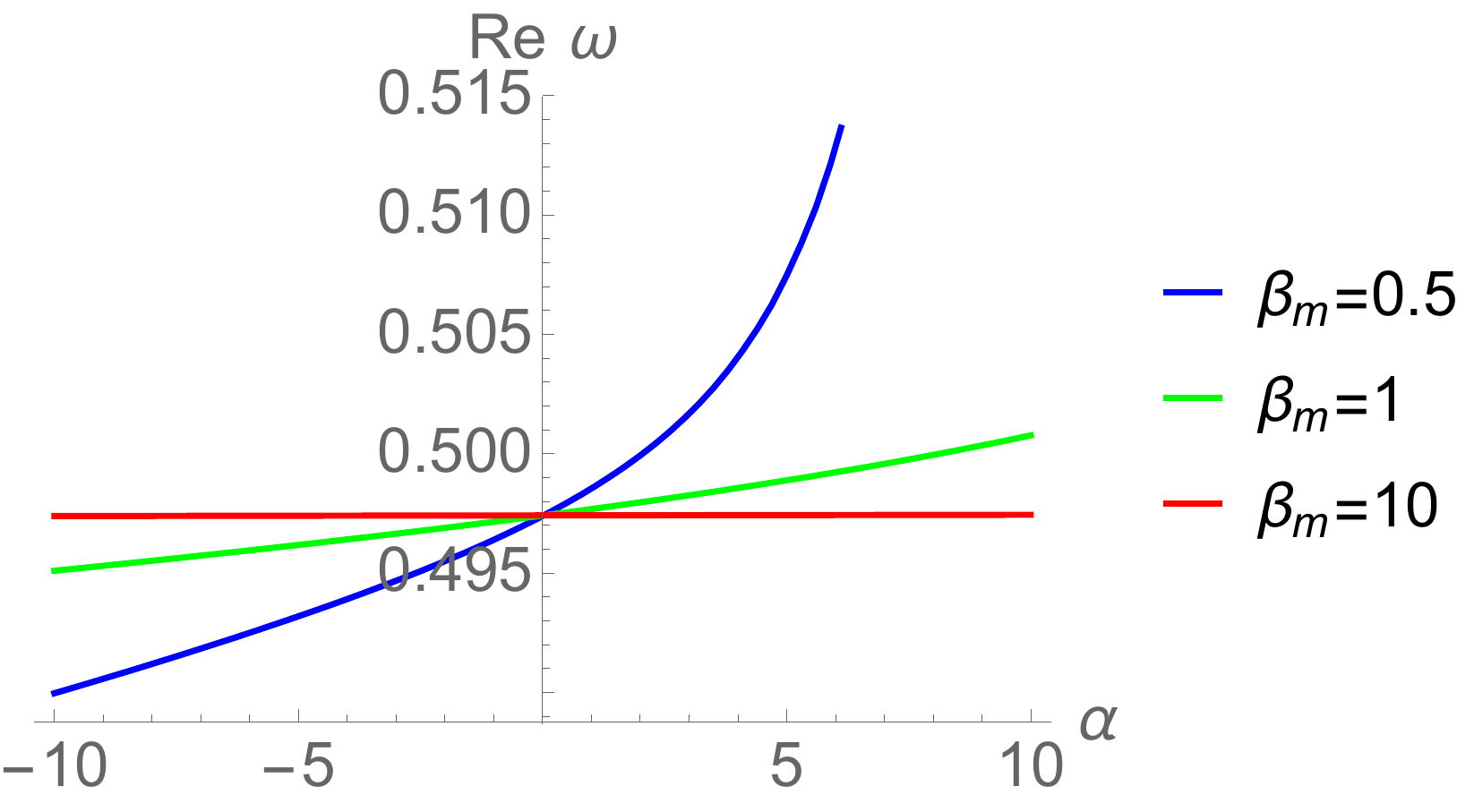}
\includegraphics[scale=0.5]{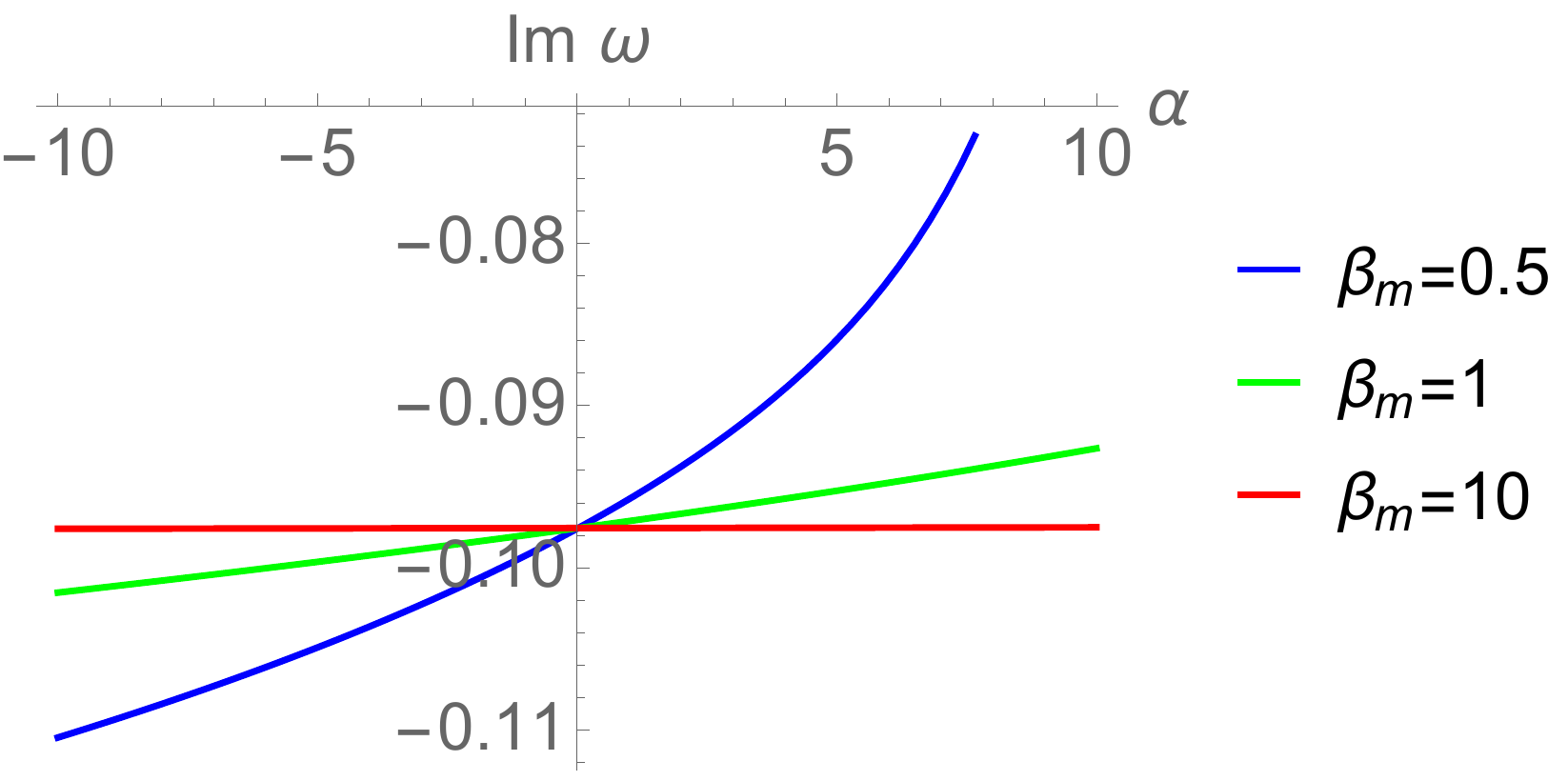}
\caption{The QNM frequencies of $R+\alpha R^2$ Born-Infeld black holes are shown in terms of $\alpha$. Each curve corresponds to different values of $\beta_m$ and we consider the fundamental mode $n=0$ and $l=2$, fixing the value of charge $Q=0.2$.} 
\label{fig4}
\end{figure}

In FIG.~\ref{fig4}, we fix the charge $Q=0.2$ and the multiple $l=2$, then present the fundamental QNM frequencies in terms of $\alpha$. Different colors of curves correspond to different values of Born-Infeld constant $\beta_m$. 

It can be seen that the real part of the frequency and the decay time would increase when $\alpha$ increases. Furthermore, even though the deviations of the QNM resulting from the Born-Infeld corrections are very small when $\alpha=0$, these deviations can be significantly amplified by changing $\alpha$, as long as $\beta_m$ is small enough. If $\beta_m$ is large enough (for instance, see the red curves in FIG.~\ref{fig4}), the QNM frequency remains almost the same when changing $\alpha$. This is also expected because when $\beta_m$ is large, the Born-Infeld NED reduces to the standard Maxwell field and the Palatini $R+\alpha R^2$ theory reduces to GR without a cosmological constant. The QNM frequencies, therefore, will be close to those of the standard RN charged black hole, independent of the value of $\alpha$. 

\section{Conclusion}\label{conclu}
In this paper, we investigate the QNM frequencies of the massless scalar perturbations to various charged black holes, which are based on different Palatini-type gravity theories. We first consider the Palatini $f(R)$ gravity coupled with Born-Infeld NED and pay more particular attention to the cases where $f(R)=R$ (the Einstein-Born-Infeld black hole) and $f(R)=R\pm R^2$, respectively. Second, we consider charged black holes within the EiBI gravity coupled with linear electromagnetic fields. Both positive and negative signs of the Born-Infeld coupling constant are taken into account in the EiBI model. The QNM frequencies are calculated with the WKB method up to the 6th order, which is believed to be accurate for the modes whose multipole number is larger than the overtone $l>n$. These modes are important from astrophysical points of view because they correspond to a longer decay time and would dominate the late time ringdown signals. We find that the black holes considered here are all stable against the massless scalar perturbations and their QNM frequencies would deviate from those of the RN black hole due to the Born-Infeld modifications. In particular, one can distinguish contributions to the QNM frequencies between that from the gravity sector (EiBI) and that from the matter sector (Einstein-BI). Furthermore, the QNMs in the eikonal limit $(l\rightarrow\infty)$ are studied based on their correspondence with the properties of the null circular orbit around the black hole. The qualitative behaviors of the QNM frequencies are almost the same as the cases with smaller $l$.

In the comparison between the Einstein-Born-Infeld black hole and the EiBI charged black hole, we find that the oscillation frequencies and the decay time of the perturbations increase with $1/\beta_g$ for the EiBI charged black hole with a positive Born-Infeld coupling constant ($\epsilon>0$). If $\epsilon<0$, the deviations of the frequencies from the standard RN black hole have opposite behaviors. On the other hand, we find that the Born-Infeld modifications $1/\beta_m$ from the matter part of the action would decrease the oscillation frequencies as well as the decay time of the QNMs for the Einstein-Born-Infeld black hole (as long as $\beta_m$ is large enough). Furthermore, the deviations of the Einstein-Born-Infeld black hole are less sensitive than those of the EiBI charged black hole to the changes of $1/\beta$. This means that for the same values of $\beta_m$ and $\beta_g$, the deviations of the QNMs of the EiBI charged black holes from the RN black hole are usually larger.  

For the $R+\alpha R^2$ gravity, we find that the oscillation frequencies and the decay time increase with $\alpha$. By including the quadratic correction $R^2$ in the action, the deviations resulting from the Born-Infeld NED can be amplified, as long as $\beta_m$ is not too large (see FIG.~\ref{fig4}). 

Since we have shown that these charged black holes are stable against the massless scalar perturbations in the parameter space of interest, it is natural and necessary to study the QNMs of the electromagnetic perturbations and gravitational perturbations. In principle this is not an easy task since the electromagnetic fields and the gravitational fields are coupled with each other in the case of charged black holes. But the QNMs of these fields are by nature more related to the real ringdown signals emitted by gravitational waves. Furthermore, one can also consider the theory with non-minimal couplings between the matter Lagrangian and gravity. The non-minimal couplings would violate the standard conservation equation of the energy momentum tensor, hence the Klein-Gordon equation \eqref{KG} should be modified. We expect that with the further improvement of the precision of future gravitational wave astronomy, one can get a much better understanding and a more stringent constraint on these different types of gravitational theories. We leave these interesting issues for our coming works.

\acknowledgments

CYC would like to thank R. A. Konoplya for providing the WKB approximation. CYC and PC are supported by Taiwan National Science Council under Project No. NSC 97-2112-M-002-026-MY3, Leung Center for Cosmology and Particle Astrophysics (LeCosPA) of National Taiwan University, and Taiwan National Center for Theoretical Sciences (NCTS). PC is in addition supported by US Department of Energy under Contract No. DE-AC03-76SF00515.

\end{document}